\documentclass[12pt]{article}
\pdfoutput=1
\usepackage{latexsym}
\usepackage{cite,epsfig,amssymb,euscript,slashed}
\usepackage[normalem]{ulem} %defines \sout{word} which strikes out {word}
\usepackage{amsmath}
\usepackage{float}
\usepackage[linktocpage]{hyperref}
\usepackage{bbm}
\usepackage{braket}
\usepackage{xcolor}
\usepackage{fancybox}
\usepackage{geometry}
\if{}
\fi

\def\be{\begin{equation}}
\def\ee{\end{equation}}
\def\bseq{\begin{subequations}}
\def\eseq{\end{subequations}}

\def\bea{\begin{eqnarray}}
\def\eea{\end{eqnarray}}

\def\bseq{\begin{subequations}}
\def\eseq{\end{subequations}}

\numberwithin{equation}{section} %%
\usepackage[]{graphicx}

\def\ii           {{\rm i}}

\def\sqr#1#2{{\vcenter{\vbox{\hrule height.#2pt
 \hbox{\vrule width.#2pt height#1pt \kern#1pt \vrule width.#2pt}\hrule
 height.#2pt}}}}

\def\slashchar#1{\setbox0=\hbox{$#1$}           % set a box for #1
\dimen0=\wd0                                 % and get its size
\setbox1=\hbox{/} \dimen1=\wd1               % get siste of /
\ifdim\dimen0>\dimen1                        % #1 is bigger
\rlap{\hbox to \dimen0{\hfil/\hfil}}      % so center / in box
#1                                        % and print #1
\else                                        % / is bigger
\rlap{\hbox to \dimen1{\hfil$#1$\hfil}}   % so center #1
/                                         % and print /
\fi}

\begin{document}
\font\cmss=cmss10 \font\cmsss=cmss10 at 7pt

\title{Introductory Notes on Non-linear Electrodynamics \linebreak
and its Applications}

\author{ Dmitri P. Sorokin \footnote{e-mail: {\tt  dmitri.sorokin@pd.infn.it }}}

\date{}

\maketitle

\vspace{-1.5cm}

\begin{center}

\vspace{0.5cm}
\textit{I.N.F.N. Sezione di Padova, Via F. Marzolo 8, 35131 Padova, Italy\\
and
\\
Dipartimento di Fisica e Astronomia ``Galileo Galilei",  Universit\`a degli Studi di Padova
}
\end{center}

\vspace{5pt}

\abstract{In 1933-1934 Born and Infeld constructed the first non-linear generalization of Maxwell's electrodynamics that turned out to be a remarkable theory in many respects. In 1935 Heisenberg and Euler computed a complete effective action describing non-linear corrections to Maxwell's theory due to quantum electron-positron one-loop effects. Since then, these and a variety of other models of non-linear electrodynamics proposed in the course of decades have been extensively studied and used in a wide range of areas of theoretical physics including string theory, gravity, cosmology and condensed matter (CMT). In these notes I will overview general properties of non-linear electrodynamics and particular models which are distinguished by their symmetries and physical properties, such as a recently discovered unique non-linear modification of Maxwell's electrodynamics which is conformal and duality invariant. I will also sketch how non-linear electromagnetic effects may manifest themselves in physical phenomena (such as vacuum birefringence), in properties of gravitational objects (e.g. charged black holes) and in the evolution of the universe, and can be used, via gravity/CMT holography, for the description of properties of certain conducting and insulating materials.
}
\noindent

%\noindent {\em Possible comment ............  }

\thispagestyle{empty}

%\vfill
%\vskip 5.mm
%\hrule width 5.cm
%\vskip 2.mm
%{\scriptsize
%\noindent e-mails: {\tt sorokin@pd.infn.it
%}}

\newpage

\setcounter{footnote}{0}

\tableofcontents

\newpage

\section{Preface}
Models of non-linear electrodynamics  (NED for short) have been extensively studied as possible guides to catch new physics: e.g. for tackling fundamental cosmological problems (such as physics of black holes, inflation and dark energy), and for an effective description of properties of certain condensed matter materials and optical media.
Notable examples are the famous Born-Infeld theory \cite{Born:1934gh} (based on earlier work of Born \cite{Born:1933qff,Born:1934ji}), which was invented to ensure that electric field self-energy of charged point particles is finite but much later turned out to be an important ingredient of String Theory (see e.g. \cite{Gibbons:2001gy} for a review), and the Heisenberg-Euler effective theory of quantum electrodynamics \cite{Heisenberg:1936nmg} (see \cite{Dunne:2012vv} for a historical review).

In addition to Born-Infeld and Heisenberg-Euler theories, there exists a variety of models of non-linear electrodynamics containing such instances as logarithmic, double-logarithmic, exponential, rational, arcsin, power-law, inverse, quasitopological and other NEDs (see e.g.  \cite{Soleng:1995kn,Ayon-Beato:1998hmi,Bronnikov:2000vy,Ayon-Beato:2000mjt,Cataldo:2000we,Burinskii:2002pz,Hassaine:2007py,Jing:2011vz,Hendi:2012zz,Cembranos:2014hwa,Gaete:2014nda,Kruglov:2014hpa,Kruglov:2016ezw,Kruglov:2017fck,Liu:2019rib,Gullu:2020ant,Balakin:2021jby,Balakin:2021arf,Dehghani:2021fwb}). They have been created for applications in gravity and cosmology, as well as for a gravity/CMT holographic description of certain strongly-coupled condensed matter systems. The newest recently found specimen in this ``NED zoo" is ModMax, the unique non-linear modification of Maxwell’s theory that preserves all its symmetries including electric-magnetic duality and conformal invariance \cite{Bandos:2020jsw}. We came upon this finding somewhat occasionally when studying conformal limits of Born-Infeld electrodynamics and its higher-rank gauge field generalizations on branes in the context of string and M-theory \cite{Bandos:2020hgy}.

In these notes I will overview basic properties of generic NEDs and the most distinguished examples of NEDs in four-dimensional space-time including Heisenberg-Euler, Born-Infeld, Bialynicki-Birula \cite{BialynickiBirula:1983tx,BialynickiBirula:1992qj} and ModMax, and will flash some effects via which non-linearities of these and other models may manifest themselves in physical phenomena, such as vacuum birefringence, and their role in the context of gravity and condensed matter theory.
I would like to keep the presentation as simple as possible and refer the reader for more details to the original literature and more extended reviews listed in a very incomplete Bibliography.

\section{Maxwell's electrodynamics}
To set the stage and introduce the notation and main ingredients, let us start with the very well known example of free Maxwell's theory in four-dimensional space-time. Its action can be written in the manifestly Lorentz-invariant and gauge-invariant form with the use of the field strength $F_{\mu\nu}=\partial_\mu A_\mu-\partial_\mu A_\mu$ of the four-vector potential $A_\mu(x)$, or in the non-manifestly Lorentz invariant form in terms of the electric and magnetic three-vector fields $\mathbf E$ and $\mathbf B$, which are components of $F_{\mu\nu}$ \footnote{In the most of the expressions we use natural units in which the speed of light $c$, the vacuum permittivity $\varepsilon_0$ and the Plank constant $\hbar$ are set to one.}:
\be\label{MaxwellS}
\mathcal S=-\frac 14\int d^4 x \, F_{\mu\nu}F^{\mu\nu}= \frac 12 \int d^4 x \,({\mathbf E}^2-{\mathbf B}^2)
\ee
$$
E_i=\partial_0 A_i - \partial_i A_0, \qquad B^i=\frac 12 \varepsilon^{ijk}F_{jk},
\qquad
\mu,\nu=(0,i);\qquad i,j,k=1,2,3\,.
$$
Maxwell's equations, including the $F_{\mu\nu}$ Bianchi identities, are
\bea\label{ME1}
\partial_\mu F^{\mu\nu}&=& 0,\\
\partial_\mu \tilde F^{\mu\nu} &=& 0\, ,\qquad \qquad \tilde F^{\mu\nu}=\frac 12\varepsilon^{\mu\nu\rho\sigma}F_{\rho\sigma},\nonumber
\eea
or
\bea\label{ME2}
&\frac{\partial{\mathbf E}}{\partial t}=\nabla \times {\mathbf B}, \qquad \nabla \cdot{\mathbf E=0},\\
&\frac{\partial{\mathbf B}}{\partial t}=-\nabla \times {\mathbf E}, \qquad \nabla \cdot{\mathbf B=0}.\nonumber
\eea
As is well known, the free Maxwell equations are invariant under the $SO(2)$-duality rotations of $F_{\mu\nu}$ and its Hodge dual $\tilde F_{\mu\nu}$ (or $\mathbf E$ and $\mathbf B$)
$$
 \begin{pmatrix}F'^{\,\mu\nu}\\
 \tilde F'^{\,\mu\nu}\end{pmatrix}
=
\begin{pmatrix} \cos\alpha & \sin\alpha\\
 -\sin\alpha & \cos\alpha \end{pmatrix}\,
\begin{pmatrix}F^{\mu\nu}\\
 \tilde F^{\mu\nu}\end{pmatrix}\,.
$$
The Maxwell action is not duality-invariant, but there exist formulations in which the duality invariance gets lifted off-shell, i.e. to the level of the action. We will not discuss these manifestly duality-symmetric actions in these notes.

However, as is well known, the Maxwell action is invariant under a larger, conformal symmetry which includes the Poincar\'e group, dilatations and special conformal transformations. For simplicity, let me only present the dilatation (re-scaling) of the space-time coordinates and the gauge vector potential with a constant parameter $a$ which leaves the Maxwell action invariant
\be\label{dilatation}
x^\mu \to a\, x^\mu, \quad A_\mu \to a^{-1} A_\mu, \quad F_{\mu\nu} \to a^{-2} F_{\mu\nu}.
\ee

Note that conformal symmetry and duality invariance play an important role in various areas of theoretical physics, from condensed matter to the theory of fundamental interactions, and in particular in String Theory. These fundamental symmetries are among basic principals assumed to be satisfied by a variety of theoretical models at least in certain limits in which these symmetries are expected. It is therefore natural to consider implications of conformal symmetry and duality invariance also in the context of non-linear generalizations of Maxwell's electrodynamics.

\section{Non-linear electrodynamics}
The Maxwell action is quadratic in $F_{\mu\nu}$ and Maxwell's equations are linear. But Maxwell's electrodynamics may receive non-linear corrections of higher orders in $F_{\mu\nu}$ and may thus be extended to a non-linear theory. The nature and form of these corrections can be different, depending on a model.

\subsection{Heisenberg-Euler-Kockel theory}

For instance, in the Heisenberg-Euler theory \cite{Heisenberg:1936nmg} the higher order corrections appear as a result of the one-loop effects produced by virtual electrons and positrons circulating in the quantum loops. More precisely the Heisenberg-Euler action is the full non-perturbative one-loop effective action for quantum spinor electrodynamics in (all orders of) a uniform background electromagnetic field and describes QED vacuum polarization effects. Its Lagrangian density is
\bea\label{EHL}
\mathcal L_{HE}&=&\frac 12 (\mathbf E^2-\mathbf B^2)\\
&+&\frac{2\pi e^2}{\hbar c}\int_0^\infty\, \frac{d\eta}{\eta^3} e^{-\eta}\left\{{\tt i}\eta^2 (\mathbf E\cdot\mathbf B)\,\frac{\cos\left(\frac\eta{\mathcal E_c}\sqrt{\mathbf E^2-\mathbf B^2+2{\ii} (\mathbf E\cdot\mathbf B)}\right)+ c.c}{\cos\left(\frac\eta{\mathcal E_c}\sqrt{\mathbf E^2-\mathbf B^2+2{\ii} (\mathbf E\cdot\mathbf B)}\right)- c.c}+\mathcal E^2_c-\frac{\eta^3}3(\mathbf E^2-\mathbf B^2)\right\}\,,\nonumber
\eea
 where the presence of the speed of light $c$ and the Planck constant $\hbar$ has been made explicit, and $\mathcal E_c=\frac{m_e^2c^3}{e\hbar } \sim 10^{16} V/{\tt cm}$ (with $m_e$ being the electron mass) is the critical electric field strength above which the creation of physical electron-positron pairs from vacuum occurs.

 Up to the fourth order in $F_{\mu\nu}$ the Heisenberg-Euler Lagrangian density reduces to the effective Lagrangian density computed earlier (in 1935) for low energy photons $\hbar \omega \ll m_e c^2$ by Euler and Kockel \cite{Euler:1935zz}
\bea\label{EKL}
\mathcal L_{spinor\, QED}&=&-\frac 1{4\mu_0} F_{\mu\nu}F^{\mu\nu}+\frac{\hbar^3\alpha^2}{90 \mu_0^2 m_e^4 c^5}\left((F_{\mu\nu}F^{\mu\nu})^2+\frac 74(F_{\mu\nu}\tilde F^{\mu\nu})^2\right)+\mathcal O\left((FF)^3\right)
\nonumber\\
&=&\frac 12 \,({\mathbf E}^2-{\mathbf B}^2)+\frac{2\hbar^3\alpha^2}{45\mu_0^2 m_e^4c^5}\left(({\mathbf E}^2-{\mathbf B}^2)^2+ 7 (\mathbf E\cdot\mathbf B)^2\right)+\,\cdots,
\eea
where $\alpha=\frac 1{4\pi \varepsilon_0}\frac{e^2}{\hbar c}$ is the fine structure constant and $\mu_0$ is the vacuum magnetic impermeability. The coefficient of the fourth-order quantum corrections in this Lagrangian are very small
$$\frac{2\hbar^3\alpha^2}{45\mu_0 m_e^4c^5} = 1.32 \times 10^{-24}\, ({\rm Tesla}^{-2})\,.$$
This is why their effects are very hard to measure experimentally.

For comparison with the spinor electrodynamics, let us also look at the form of the effective Lagrangian density describing one-loop contributions of a massive charged scalar field to the forth order in $F_{\mu\nu}$ \cite{Weisskopf:1936hya,Schwinger:1951nm}
\bea\label{SED}
\mathcal L_{scalar\, QED}&=&-\frac 1{4\mu_0} F_{\mu\nu}F^{\mu\nu}+\frac{\hbar^3\alpha^2}{1440 \mu_0^2 m_s^4c^5}\left(7(F_{\mu\nu}F^{\mu\nu})^2+(F_{\mu\nu}\tilde F^{\mu\nu})^2\right)+\mathcal O\left((FF)^3\right)
.
\eea

\subsection{General structure of NEDs}
The most general Lorentz-invariant action for non-linear electrodynamics in $D=4$, which usually does not include derivatives of $F_{\mu\nu}$ to avoid ghosts, is a functional of two independent Lorentz invariants, a scalar and a pseudo-scalar quadratic in $F_{\mu\nu}$ \cite{Peres:1961zz,Plebanski:1970zz}
\be\label{SP}
S=-\frac 14 F_{\mu\nu}F^{\mu\nu}=\frac 12 \,({\mathbf E}^2-{\mathbf B}^2)\,,\qquad P=-\frac 14 F_{\mu\nu}\tilde F^{\mu\nu}={\mathbf E}\cdot{\mathbf B}\,.
\ee
So the generic NED action is
\be\label{NEDS}
I_{NED}=\int d^4x\, \mathcal L(S,P)\,.
\ee
Usually, one assumes that $\mathcal L(S,P)$ is an analytic function at $S=P=0$ and that in the limit of weak electromagnetic fields the action reduces to that of the Maxwell theory. But, as we will see there are interesting models of non-linear electrodynamics (e.g. the conformal NEDs and in particular ModMax) which do not obey these assumptions.

Other restrictions on the form of the NED action are imposed by the physical requirements of causality and unitarity of the theory (see e.g. \cite{Gibbons:2000xe,Shabad:2009ky,Shabad:2011hf,Denisov:2017qou,Bandos:2021rqy}). These are related to the requirement that the Lagrangian density is a convex function of the electric field strength $E^i$, i.e. that the Hessian $3 \times 3$ matrix
\be\label{Hessian}
\mathcal L_{ij}=\frac{\partial^2 \mathcal L}{\partial E^i\partial E^j}
\ee
has only non-negative eigenvalues for all values of $\mathbf E$ and $\mathbf B$. The convexity of the Lagrangian density also insures the possibility of performing an involutive Legendre transform to the Hamiltonian formulation of the theory.

In contrast to the Maxwell theory a generic NED is neither conformal, nor duality-invariant, as is the Heisenberg-Euler theory. However, the effective Heisenberg-Euler-like Lagrangian for a supersymmetric system of minimally coupled one spin-1/2 and two spin-0 charged particles of equal mass is duality invariant (at least) to the fourth order in $F_{\mu\nu}$ \cite{Duff:1979bk} and coincides to this order with the Born-Infled Lagrangian (see eq. \eqref{BI} below) for an appropriate choice of its coupling constant. The fourth order terms in the supersymmetric extension of the Heisenberg-Euler Lagrangian are the sum of the forth order terms in \eqref{EKL} and twice the fourth order terms in \eqref{SED}.

If one is interested in maintaining the conformal invariance, than the Lagrangian density of the non-linear electrodynamics must be a homogeneous function of $S$ and $P$ of degree one under the scale transformations \eqref{dilatation} of the coordinates and the gauge vector potential $A_\mu$
\be\label{ConfL}
\mathcal L(S,P) \quad \to \quad \mathcal L(a^{-4} S, a^{-4} P)= a^{-4}\mathcal L(S,P).
\ee
Then in the action the scaling factor of the Lagrangian density gets compensated by the opposite re-scaling of the integration measure.
Various conformal non-linear electrodynamics models and their applications (mainly to black holes but also in the AdS/CFT contexts) have been studied (in different space-time dimensions) for more than two decades (see e.g.  \cite{Cataldo:2000we,Hassaine:2007py,Jing:2011vz,Cembranos:2014hwa,Denisov:2017qou,Denisova:2019lgr,Duplij:2019jty,Asorey:2021rwv,Kokoska:2021lrn}).

As far as the electric-magnetic duality invariance of the field equations of non-linear electrodynamics is concerned, the condition which this invariance imposes on the form of the Lagrangian density was first derived by Bialynicki-Birula in 1983 \cite{BialynickiBirula:1983tx}. It has the following form
\be\label{DINED}
F_{\mu\nu} \tilde F^{\mu\nu}-2\varepsilon^{\mu\nu\lambda\rho}\, \frac{\partial\mathcal L}{\partial F^{\mu\nu}}\,\frac{\partial\mathcal L}{\partial F^{\lambda\rho}}=0 \qquad \Rightarrow \qquad {\mathcal L}_S^2 - \frac{2S}{P}{\mathcal L}_S {\cal L}_P - {\mathcal L}_P^2 =1,
\ee
where $\mathcal L_S= \frac{\partial\mathcal L}{\partial S}$ and $\mathcal L_P= \frac{\partial\mathcal L}{\partial P}$.

When the condition \eqref{DINED} holds, the Euler-Lagrange equations and the Bianchi identities of the NED
\begin{eqnarray}
&\partial_\mu \left(-2\frac{\partial\mathcal L}{\partial F_{\mu\nu}}\right)=0, \\
&\partial_\mu \tilde F^{\mu\nu}=0\nonumber
\end{eqnarray}
are transformed into each other by the SO(2)-duality rotations
\be\label{DR}
\begin{pmatrix}-2\frac{{\partial\mathcal L(F')}}{\partial F^\prime_{\mu\nu}}\\
 \tilde F'^{\mu\nu}\end{pmatrix}
=
\begin{pmatrix} \cos\alpha & \sin\alpha\\
 -\sin\alpha & \cos\alpha \end{pmatrix}\,
\begin{pmatrix}-2\frac{\partial\mathcal L}{\partial F_{\mu\nu}}\\
 \tilde F^{\mu\nu}\end{pmatrix}\,,
\ee
where $\mathcal L(F')$ has the same form as $\mathcal L(F)$. This, of course, does not mean that the NED Lagrangian is invariant under the duality rotations \eqref{DR}. In fact, it does not but the corresponding energy-momentum tensor \cite{Gaillard:1981rj} and the Hamiltonian \cite{BialynickiBirula:1983tx} are duality-invariant.

In the paper of 1983 Bialynicki-Birula showed that the Lagrangian density of the Born-Infeld theory \eqref{BI} satisfies the condition \eqref{DINED} and hence that this theory is duality invariant, although probably the first evidence of the duality invariance of the Born-Infeld theory appeared in a paper by Schr\"odinger of 1935 \cite{Schrodinger:1935oqa}. The properties of the duality-invariance condition \eqref{DINED} its (general) solutions and an $SL(2,R)$-duality extension to the case of NEDs coupled to an axion-dilaton were further analyzed in detail in \cite{Gibbons:1995cv,Gibbons:1995ap} and \cite{Gaillard:1997zr,Gaillard:1997rt}, the latter papers having been the development of earlier work \cite{Gaillard:1981rj} of Gaillard and Zumino on the general theory of duality-invariant nonlinear electrodynamics. More explicit examples of duality-invariant NED Lagrangians were found in \cite{Hatsuda:1999ys}. A generalization of this condition to the case of electrodynamics non-minimally coupled to a higher-curvature gravity was recently considered in \cite{Cano:2021tfs,Cano:2021hje}.

\section{Born-Infeld theory and its weak and strong field limits}
The Born-Infeld Lagrangian density has the following form
\begin{eqnarray}\label{BI}
\mathcal L_{BI}&=&T-T\sqrt{-\det (\eta_{\mu\nu}+T^{-\frac 12}F_{\mu\nu})}=T-\sqrt{T^2+\frac T2 F_{\mu\nu}F^{\mu\nu}-\frac 1{16}(F_{\mu\nu}\tilde F^{\mu\nu})^2}\, \nonumber\\
&=& - \frac 14\,F_{\mu\nu}F^{\mu\nu}+\frac 1{32\,T}\left((F_{\mu\nu}F^{\mu\nu})^2+(F_{\mu\nu}\tilde F^{\mu\nu})^2\right)+\mathcal O\left((F^2)^3\right)\,,
\end{eqnarray}
where $T$ is the coupling parameter of the dimension of energy density.

The original motivation of Born and Infeld for constructing this theory was to ensure finiteness of the electric field self-energy of charged particles, which is guaranteed by the condition that the expression under the square root may only have non-negative values. Much later it was discovered that the Born-Infeld theory is an important ingredient of String Theory.
In bosonic open string theory the Born-Infeld action arises as an effective low-energy action \cite{Fradkin:1985qd}, while in type II superstring theories it describes an effective worldvolume theory of Dirichlet branes (in ten-dimensional space-time) on which open strings end \cite{Leigh:1989jq}, and $T$ has the meaning of brane tension. Since $T$ is dimensional, the Born-Infeld theory is not conformal, but it is duality-invariant, as we have mentioned above.

\subsection{Conformal and duality-invariant non-linear electrodynamics of Bialynicki-Birula}

 As is well known and not hard to see from \eqref{BI}, in the infinite tension limit $T\to\infty$ (which is equivalent to the weak electromagnetic field limit), the Born-Infeld Lagrangian density reduces to that of Maxwell's electrodynamics \eqref{MaxwellS}. A much less known limiting theory is the one appearing in the strong field limit of Born-Infeld theory \cite{BialynickiBirula:1983tx,BialynickiBirula:1992qj}, which is equivalent to take $T\to 0$. A reason why this limit is less known is that it cannot be taken in the Lagrangian formulation. Indeed, in this limit the Born-Infeld Lagrangian density becomes a purely imaginary total derivative
 $$
 \mathcal L_{BI}|_{T\to 0}=\frac {\tt i}4|F_{\mu\nu}\tilde F^{\mu\nu}|\,,
 $$
 hence the corresponding action vanishes (modulo an un-physical imaginary boundary Chern-Simons term).

 To take the strong field limit it is necessary to pass (via the Legendre transform) to the Hamiltonian formulation of the Born-Infeld theory as was noticed by Bialynicki-Birula in 1983 \cite{BialynickiBirula:1983tx,BialynickiBirula:1992qj}. The Born-Infeld Hamiltonian density is
 \be\label{HBI}
 \mathcal H_{BI}(\mathbf D, \mathbf B)=\mathbf D\cdot \mathbf E- \mathcal L_{BI}(\mathbf E, \mathbf B)=\sqrt{T^2 + T ( {\bf D}^2 + {\bf B}^2) + |{\bf D}\times {\bf B}|^2} -T\,,
 \ee
where $\mathbf D=\frac{\partial\mathcal L}{\partial \mathbf E}$ is the electric displacement vector field playing the role of the canonical momentum conjugate to the three-vector gauge potential $\mathbf A$.

Now the tensionless (strong field) limit $T\to 0$ can be taken and one gets the Hamiltonian density first obtained by Bialynicki-Birula in 1983
\be\label{BBH}
\mathcal H_{BB}=|{\bf D}\times {\bf B}|\,.
\ee
As Bialynicki-Birula discussed in detail in \cite{BialynickiBirula:1992qj}, the theory described by this Hamiltonian is rather remarkable. It is invariant under $SL(2,R)$-duality (rather than SO(2)), which can be immediately seen from the invariance of the Hamiltonian density \eqref{BBH} under the $SL(2,R)$ transformations of $\mathbf B^I=(\mathbf D,\mathbf B)$ ($\mathbf B'^I=L^I{}_{J}\,\mathbf B^J, \,\,\det L^I{}_J=1, \,\,I,J=1,2$). It is conformal and possesses infinite number of conserved currents with the associated tensorial conserved charges
$$
Q^{i_1i_2\cdots i_k}=\int dx^3\, \mathcal H_{BB}\,n^{i_1}n^{i_2}\cdots n^{i_k}, \qquad \mathbf n=\frac {\mathbf D\times \mathbf B}{|\mathbf D\times \mathbf B|},
$$
which indicates that this theory can be integrable. As is seen from the form of its Hamiltonian, the Bialynicki-Birula theory does not include Maxwell's electrodynamics as a whole. It can be shown that this theory describes all possible configurations of the so-called null electromagnetic fields whose scalar and pseudo-scalar Lorentz invariants \eqref{SP} vanish
\be\label{nullfields}
S=-\frac 14 F_{\mu\nu}F^{\mu\nu}=\frac 12(\mathbf E^2-\mathbf B^2)=0, \qquad P=-\frac 14 F_{\mu\nu}\tilde F^{\mu\nu}= \mathbf E\cdot \mathbf B=0.
\ee
The Bialyanicki-Birula theory can be associated with an infra-red limit of a D3-brane \cite{Mezincescu:2019vxk} and has its counterparts in higher space-time dimensions \cite{Bandos:2020hgy}. For instance, in $D=6$ it is a strong-field limit \cite{Townsend:2019ils,Bandos:2020hgy} of the non-linear Born-Infeld-like theory for a chiral 2-form gauge field \cite{Perry:1996mk,Pasti:1997gx}, while in $D=10$ the theory whose strong field limit is the BB-like chiral 4-form electrodynamics is still unknown \cite{Bandos:2020hgy}.

A natural question arises whether there are other conformal and duality invariant models of electrodynamics in addition to the Maxwell and Bialynicki-Birula theory? The answer is yes, there is but only one more non-linear electrodynamics with these symmetry properties called ModMax \cite{Bandos:2020jsw}.

\section{ModMax electrodynamics}
The Lagrangian density of ModMax has the following form \cite{Bandos:2020jsw,Kosyakov:2020wxv}.
\begin{eqnarray}\label{MML}
{\cal L}_{ModMax}& = -\frac {\cosh\gamma} 4\, F_{\mu\nu}F^{\mu\nu}+ \frac{\sinh\gamma}4\, \sqrt{(F_{\mu\nu}F^{\mu\nu})^2+(F_{\mu\nu}\tilde F^{\mu\nu})^2}\, \\
&\nonumber\\
&= \frac {\cosh\gamma} 2\, (\mathbf E^2-\mathbf B^2)+ \frac{\sinh\gamma}2\, \sqrt{(\mathbf E^2-\mathbf B^2)^2+4(\mathbf E\cdot \mathbf B)^2}\,,\nonumber
\end{eqnarray}
where $\gamma$ is the dimensionless coupling constant. The conditions of causality and unitarity require this constant be non-negative $\gamma \geq 0$. These values of $\gamma$ also ensure that the Lagrangian density is a convex function of the electric field $E^i$ \cite{Bandos:2020jsw} and that its energy-momentum tensor satisfies the weak, strong and dominant energy conditions \cite{Pergola:2021}. Note that Maxwell's electrodynamics is not a weak field limit of ModMax because of conformal invariance, but is recovered when the ModMax coupling constant tends to zero $\gamma\to 0$.

The field equations of the theory are
\begin{equation}\label{Gtilda}
\cosh\gamma\,\partial_\mu F^{\mu\nu} +\sinh\gamma\,\partial_\mu\,\Bigl(\frac{SF^{\mu\nu} + P\Tilde{F}^{\mu\nu}}{\sqrt{S^{2} + P^{2}}}\Bigr)=0.
\end{equation}
They are non-linear, but linearize by field configurations for which $P=a\,S$ with $a$ being a constant. So all the solutions of Maxwell's equations with $P=a\,S$ are solutions of ModMax theory.

Since ModMax is conformal, its energy-momentum tensor is traceless and proportional to that of Maxwell's theory
\begin{eqnarray}\label{EMT}
T^{\mu\nu}&=&\left(F^{\mu}{}_{\rho}F^{\nu\rho}-\frac 14 \eta^{\mu\nu} (F_{\rho\lambda}F^{\rho\lambda})\right){\mathcal L_S}\,,\\
\mathcal L_S&=&\frac{\partial\mathcal L}{\partial S}=\left(\cosh\gamma-\sinh\gamma\frac{F_{\sigma\nu}F^{\sigma\nu}}{\sqrt{(FF)^2+(F\tilde F)^2}}\right)\,,\nonumber\\
T^{\mu}{}_{\mu}&=0\,.&\nonumber
\end{eqnarray}
One can see that the energy-momentum tensor is non-analytic and is not well defined when the electromagnetic fields are null \eqref{nullfields}. The ambiguity of the value of the scalar factor $\mathcal L_S$ of $T^{\mu\nu}$ in the null-field limit ranges from $e^{-\gamma}$ to $e^\gamma$. This might be an issue for the theory, since the class of solutions for which the electromagnetic fields are null, such as the plane waves, are not well defined in the ModMax Lagrangian formulation.\footnote{Note, though, that the vacuum (in which $F_{\mu\nu}=0$) is a well defined solution of the ModMax Lagrangian field equations \eqref{Gtilda} with zero energy-momentum tensor.} However, somewhat surprisingly, the Hamiltonian formulation of ModMax comes to the rescue \cite{Bandos:2020jsw}. The ModMax Hamiltonian density is obtained from its Lagrangian density by the Legendre transform and has the following form \footnote{A detailed analysis of the ModMax Hamiltonian formulation in the phase space associated with the Plebanski variables \cite{Plebanski:1970zz} has been carried out in \cite{Escobar:2021mpx}.}
\be\label{MMH}
\mathcal H_{MM}=\frac 12 \left(\cosh\gamma\,(\mathbf D^2+\mathbf B^2)-\sinh\gamma \sqrt{(\mathbf D^2+\mathbf B^2)^2-4(\mathbf D\times \mathbf B)^2}\right)+A_0\,\boldsymbol{\nabla} \cdot \mathbf D \,,
\ee
where in the last term the time component of the $4D$ gauge vector field plays the role of the Lagrange multiplier that insures the Gauss law constraint $\boldsymbol{\nabla}\cdot\mathbf D=0$.

Note that the Hamiltonian density is the phase-space counterpart of the energy density
\be\label{T00}
T^{00}=\frac 1{2} \left({\mathbf E}^2+{\mathbf B}^2\right)\left(\cosh\gamma+\sinh\gamma\frac{{\mathbf E}^2-{\mathbf B}^2}{\sqrt{({\mathbf E}^2-{\mathbf B}^2)^2+4({\mathbf E}\cdot {\mathbf B})^2}}\right)\geq 0\,.
\ee
However, in contrast to the Lagrangian energy density, $\mathcal H_{MM}$ is well defined also for the null-field configurations \eqref{nullfields} for which the electric displacement field $\mathbf D=\frac{\partial \mathcal L}{\partial \mathbf E}$ and the magnetic field $\mathbf B$ satisfy the following condition
\be\label{nullDB}
\mathbf D^2+\mathbf B^2=2 \cosh\gamma\,|\mathbf D\times  \mathbf B|\,,
\ee
and the Hamiltonian density reduces to
\be\label{nullMM}
\mathcal H|_{null-fields}=|\mathbf D\times \mathbf B|\,.
\ee
As one might expect, for the null fields the ModMax Hamiltonian density coincides with that of Bialynicki-Birula \eqref{BBH}.

The condition \eqref{nullDB} is also the bound on $\mathbf D$ and $\mathbf B$ above which the ModMax Hamiltonian density (with $\gamma \geq 0$) is a convex function of $\mathbf D$ \cite{Bandos:2020jsw}. Indeed, it can be shown that the Hessian matrix $H_{ij}=\frac{\partial{\mathcal H_{MM}}}{\partial D^i\partial D^j}$ has only non-negative eigenvalues iff
\be\label{convexbound}
\mathbf D^2+\mathbf B^2\geq 2 \cosh\gamma\,|\mathbf D\times  \mathbf B|\,.
\ee
Only if the electric displacement fields $\mathbf D$ satisfy eq. \eqref{convexbound} the Legendre transform to the Lagrangian density \eqref{MML} is involutive. This raises an interesting general question how to handle theories (especially at quantum level) whose phase space geometry is subject to inequality constraints like \eqref{convexbound}?

The null field configurations saturate the convexity bound \eqref{convexbound}. Among these configurations, the plane waves \cite{Bandos:2020jsw} and topologically non-trivial knotted electromagnetic fields \cite{Dassy:2021ulu} (generalizing those of Maxwell's theory \cite{Arrayas:2017sfq}) are exact solutions of the ModMax Hamiltonian equations
\begin{eqnarray}\label{feqs}
\dot {\bf B}  &=&   -\boldsymbol{\nabla}\times {\bf E} \, , \qquad \boldsymbol{\nabla} \cdot {\bf B} =0\, , \nonumber\\
\dot {\bf D}  &=& {}\;\; \boldsymbol{\nabla} \times {\bf H} \, , \qquad  \boldsymbol{\nabla} \cdot {\bf D} =0\, ,
\end{eqnarray}
where
\begin{equation}\label{EHdefs}
{\bf E} =  \partial {\mathcal H_{MM}}/\partial {\bf D} \, ,
\qquad {\bf H} = \partial {\mathcal H_{MM}}/\partial {\bf B} \, .
 \end{equation}
 For example, a linearly polarized plane wave has the following form
 \be\label{pw}
 \mathbf E=e^{-\gamma}\mathbf D=\mathbf E_0\cos (|\mathbf k|t-\mathbf k\cdot \mathbf x)=-\frac{\mathbf k}{\mathbf |\mathbf k|}\times \mathbf B\,,
 \ee
 where $\mathbf k$ is the wave vector and $\mathbf E_0$ is the wave amplitude.\footnote{Of course, the linear superposition of two plane waves is not, in general, a solution of non-linear electrodynamics equations \cite{Schrodinger:1942rwk,Schrodinger:1944lmg}.}

 This demonstrates a non-trivial complementarity of the Hamiltonian and Lagrangian formulation of ModMax electrodynamics in which the existence of the null-field solutions is not an obvious fact. In contrast, it is straightforward to see that all theories of non-linear electrodynamics whose Lagrangian reduces to that of Maxwell's theory in the weak field limit (including Born-Infeld) do have null-field solutions, such as the plane waves \cite{Schrodinger:1935oqa,Schrodinger:1942rwk,Schrodinger:1944lmg,Plebanski:1970zz}.
This is because all the non-linear terms in the field equations obtained from the Lagrangian (like \eqref{EKL} or \eqref{BI}), which can be expanded in series of positive powers of the Lorentz invariants \eqref{SP}, vanish for the null fields.

In \cite{Bandos:2020jsw,Bandos:2020hgy} ModMax and Born-Infeld theories were combined into a duality-symmetric  generalized Born-Infeld electrodynamics whose weak field limit is ModMax and the strong field limit is the Bialynicki-Birula electrodynamics.
The Lagrangian density of this theory has the following form
\be\label{MMBI}
\mathcal L_{\gamma BI}= T - \sqrt{T^2 -2T \mathcal L_{MM} - \frac 1{16}(F_{\mu\nu}\tilde F^{\mu\nu})^2}\, ,
\ee
where $\mathcal L_{MM}$ is the ModMax Lagrangian density \eqref{MML}.

In \cite{Bandos:2020hgy} these theories were lifted to a six-dimensional space-time where corresponding new models of a non-linear chiral two-form gauge field were found, and also generalized to $2n+1$-form gauge field NEDs in $D=4n+4$ with the use of the manifestly duality-symmetric and space-time covariant PST construction of their actions \cite{Pasti:1995tn,Pasti:1997gx,Buratti:2019cbm,Buratti:2019guq}. An alternative duality-symmetric space-time covariant Lagrangian formulation of ModMax was considered in \cite{Avetisyan:2021heg}. A possibility of attributing a brane-like nature to the generalized Born-Infeld theory \eqref{MMBI} was discussed in \cite{Nastase:2021uvc}.
A four-parameter family of NEDs which includes ModMax and the generalized Born-Infeld theory was considered in \cite{Kruglov:2021bhs}. It was shown there that for certain values of parameters the electric field in the center of the charged point particle and its self-energy are finite. Another deformation of ModMax which breaks duality invariance was proposed in \cite{Sokolov:2021qfs}.
$N=1$ supersymmetric extensions of ModMax and the generalized Born-Infeld electrodynamics were constructed in \cite{Kuzenko:2021cvx,Bandos:2021rqy}, and in \cite{Kuzenko:2021qcx} ModMax has been generalized to four-dimensional duality-invariant (super)conformal non-linear models of higher-spin gauge fields. A 't Hooft anomaly in the ModMax theory has been studied in \cite{Chatzistavrakidis:2021dqg}.

\section{Other specimens of non-linear electrodynamics and their applications}
In addition to Born-Infeld-like, Heisenberg-Euler-like and the generic NED,  in the last more than two decades there has been a rather extensive activity in constructing a variety of other (heuristic) models of non-linear electrodynamics
(see e.g.  \cite{Soleng:1995kn,Ayon-Beato:1998hmi,Bronnikov:2000vy,Ayon-Beato:2000mjt,Cataldo:2000we,Burinskii:2002pz,Hassaine:2007py,Jing:2011vz,Hendi:2012zz,Cembranos:2014hwa,Gaete:2014nda,Kruglov:2014hpa,Kruglov:2016ezw,Kruglov:2017fck,Liu:2019rib,Gullu:2020ant,Balakin:2021jby,Balakin:2021arf,Dehghani:2021fwb})
and using them for applications in gravity, cosmology and CMT. For instance, NEDs coupled to gravity have been used to construct new black hole solutions and study their properties and stability (see e.g. \cite{Demianski:1986wx,Wiltshire:1988uq,deOliveira:1994in,Rasheed:1997ns,Ayon-Beato:1998hmi,Ayon-Beato:1999kuh,Ayon-Beato:1999qin,Ayon-Beato:2000mjt,Bronnikov:2000yz,Bronnikov:2000vy,Yajima:2000kw,Gibbons:2000xe,Gibbons:2001sx,Burinskii:2002pz,Moreno:2002gg,Dey:2004yt,Hendi:2010zza,Miskovic:2010ui,Ruffini:2013hia,Hendi:2015hoa,Hendi:2015pwk,Hendi:2017lgb,Gulin:2017ycu,Hendi:2017ptl} and also \cite{Liu:2019rib,Cisterna:2020rkc,Nomura:2020tpc,Bokulic:2021dtz,Cano:2021tfs,Cano:2021hje,Cisterna:2021ckn,Gao:2021kvr,Daghigh:2021psm,Bokulic:2021xom,Nomura:2021efi,An:2021plu,Mkrtchyan:2022ulc,Bokulic:2022cyk} for the latest developments and review), to create cosmological models (which might avoid initial Big-Bang singularities) \cite{Garcia-Salcedo:2000ujn,DeLorenci:2002mi}, to mimic dark energy and generate an accelerated expansion of the universe \cite{Sami:2003my,Elizalde:2003ku,Novello:2003kh,Campanelli:2007cg,Kruglov:2015fbl,Ovgun:2017iwg,Kruglov:2020xfc,Kruglov:2020axn}, as well as in the context of AdS/CFT correspondence and gravity/CMT holography (see e.g. \cite{Karch:2007pd,Cai:2008in,Myers:2010pk,Jing:2011vz,Baggioli:2016oju,Kiritsis:2016cpm,Blauvelt:2017koq,Cremonini:2018kla} and \cite{Bi:2021maw} for a recent review and more references). For the study of magnetic brane solutions in modified gravity theories coupled to NEDs see e.g. \cite{Hendi:2017uly} and references there in.

 A common setup for gravity-related applications of all these models is the coupling of a NED to General Relativity or its modifications. In the case of Einstein's gravity (in general with the cosmological constant $\Lambda$) one considers Einstein's equations sourced by the energy-momentum tensor of a certain NED and some matter
\be\label{L+NED}
R^{\mu\nu}-\frac 12 g^{\mu\nu} (R-2\Lambda)=8\pi\, (T^{\mu\nu}_{NED}+T^{\mu\nu}_{matter}),
\ee
and the NED equations of motion sourced, in general, by electric and magnetic currents
\begin{eqnarray}
&\nabla_\mu \left(-2\frac{\partial\mathcal L}{\partial F_{\mu\nu}}\right)=j^\mu_e, \\
&\nabla_\mu \tilde F^{\mu\nu}=j^\mu_m\,, \label{Bianchi}
\end{eqnarray}
where $\nabla_\mu$ is the covariant derivative containing a metric connection.

Note that in the presence of magnetic monopoles or dyons, the definition of $F_{\mu\nu}$ as the curl of the vector potential $A_\mu$ should be modified and include the contribution of the Dirac string (along which $A_\mu$ is singular) for the consistency of the ``Bianchi" identity \eqref{Bianchi} (see e.g. \cite{Blagojevic:1985sh} for a review of the quantum field theory of electric and magnetic charges).

One then looks e.g. for static black-hole solutions of these equations characterized by the metric
\be\label{BHmetric}
ds^2_{BH}=-f(r)\,dt^2+\frac 1{f(r)}\,{dr}^2+u(r)\,dM_2\,,
\ee
where $f(r)$  and $u(r)$ are functions of the radial coordinate $r$, whose forms depend on the choice of the electrodynamics model and the ansatz therein. For instance, for the spherical black holes
$$
u(r)\,dM_2=r^2(d\theta^2+\sin^2\theta\,d\varphi^2)\,,
$$
where $\theta$ and $\varphi$ are the angular coordinates of a two-sphere. In AdS/CFT applications one often
chooses
$$
u(r)\,dM_2=r^2(dx^2+dy^2)\,,
$$
where $x$ and $y$ parametrize the AdS boundary at $r \to \infty$. Of course, the above choices of the gravitational metric are not the most general ones (for a more general ansatz see e.g. a recent review \cite{Baggioli:2021xuv}). The metric \eqref{BHmetric} appears in simple examples which we will now consider.

\subsection{NEDs and exact regular electric and magnetic black holes}
As was shown by Penrose \cite{Penrose:1964wq}, the black holes formed as a result of the gravitational collapse of matter in General Relativity inevitably have curvature singularities. This indicates that close to the singularity the validity of the classical Einstein's gravity breaks down and it should be replaced by a quantum theory that should resolve the singularity. This issue raised the question of the existence of regular black holes which may arise as solutions in some effective (modified) theories of gravity. The first regular black hole metric (satisfying the weak energy condition) was constructed by Bardeen in \cite{bardeen1968non} without the notion of the underlying theory whose field equations it obeys. Only much later first exact electric and magnetic regular black hole solutions were found and studied in specific models of non-linear electrodynamics coupled to General Relativity in \cite{Ayon-Beato:1998hmi,Ayon-Beato:1999kuh,Ayon-Beato:1999qin,Ayon-Beato:2000mjt,Bronnikov:2000yz,Bronnikov:2000vy,Burinskii:2002pz}. See \cite{Bokulic:2022cyk} for a review and further refining of constraints on the possibility of removing black hole singularities with NEDs.

 In \cite{Ayon-Beato:2000mjt} the Bardeen black hole was interpreted as a magnetic monopole in a certain  gravity-coupled NED which we will consider as an example. The function $f(r)$ of the Bardeen black hole metric has the following form
\be\label{BardeenBH}
f(r)=1-\frac{Mr^2}{(r^2+Q_m^2)^{3/2}}\,,
\ee
where $M$ and $Q_m$ are integration constants which (upon the result of \cite{Ayon-Beato:2000mjt}) can be regarded, respectively, as the black hole mass and magnetic charge.

Ay\'on-Beato and Gracia \cite{Ayon-Beato:2000mjt} found that a very specific NED Lagrangian density which produces the source for the Bardeen black hole in the Einstein equations has the following form
\be\label{ABGL}
\mathcal L_{ABG}(S)=\frac {3M}{|Q_m|^3}\left(\frac{|Q_m|\sqrt{-2 S}}{1+|Q_m|\,\sqrt{-2 S}}\right)^{5/2}\,, \qquad S=-\frac 14 F_{\mu\nu}F^{\mu\nu}\,,
\ee
and the electro-magnetic field strength is that of the magnetic monopole
\be\label{mf}
\frac 12 F_{\mu\nu}dx^\mu \wedge dx^\nu=Q_m\sin\theta d\theta \wedge d\varphi.
\ee
Another example of
 a NED which gives rise to a regular magnetically charged black hole was considered in \cite{Bronnikov:2000vy}. Its Lagrangian density is
\be\label{BronnikovNED}
\mathcal L=-\frac{S}{\cosh^2\left(a|2S|^{\frac 14}\right)}\,,\qquad a=const,\qquad S=-\frac 14 F_{\mu\nu}F^{\mu\nu}\,,
\ee
and the function $f(r)$ in the black hole metric sourced by \eqref{mf} has the following form
\be\label{f}
f(r)=1-\frac{|Q_m|^{\frac 32}\left(1-\tanh\left(\frac{\sqrt{|Q_m|}}{r}\right)\right)}{ar}\,.
\ee

\subsection{ModMax effects on black holes}
Also effects of the modified Maxwell electrodynamics \eqref{MML} and its generalizations \cite{Bandos:2020hgy,Kruglov:2021bhs} on properties and thermodynamics of black holes (e.g. Taub-NUT and Reissner-Nordstr\"om, accelerated ones and others), and gravitational waves have been explored in a number of papers \cite{Flores-Alfonso:2020euz,Bordo:2020aoz,Flores-Alfonso:2020nnd,Amirabi:2020mzv,Bokulic:2021dtz,Zhang:2021qga,Bokulic:2021xom,Ali:2022yys,Kruglov:2022qag,Ortaggio:2022ixh,Barrientos:2022bzm}.  Here I will restrict the discussion only to the simplest example of the Reissner-Nordstr\"om black hole.

The Reissner-Nordstr\"om black hole sourced by the ModMax energy momentum tensor has the metric \eqref{BHmetric} with the scalar function $f(r)$ of the following form
\be\label{f(r)}
f(r)=1-\frac{2M}r+\frac{e^{-\gamma}(Q^2_e+Q^2_m)}{r^2},
\ee
where $M$ is the black hole mass and $Q_e$ and $Q_m$ are its electric and magnetic charge. The corresponding electromagnetic potential is
\be\label{A}
A_\mu dx^\mu=-e^{-\gamma}\frac {Q_e}r\,dt+Q_m\,\cos\theta\,d\varphi\,.
\ee
The black hole has two horizons
\be\label{horizons}
f(r)=0 \quad \to \quad r_{\pm}=M\pm \sqrt{M^2-(Q_e^2+Q_m^2) e^{-\gamma}} \quad \to \quad M^2\geq (Q^{2}_e+Q^2_m){e^{-\gamma}}\,,
\ee
the mass squared of the extreme black hole being $M^2=(Q^{2}_e+Q^2_m){e^{-\gamma}}$. One can see that the effect of the non-linearity of ModMax consists in ``screening" the charges by the factor of $e^{-\gamma}$ (remember that for consistency of the ModMax theory $\gamma \geq 0$). So the mass squared of the extremal ModMax Reissner-Nordstr\"om black hole can be less than the sum of the squares of its charges. This, however depends on how the physically measurable charges are defined in this theory. As was shown in \cite{Lechner:2022qhb}, the screening factor $e^{-\gamma}$ can be removed from eqs. \eqref{f(r)} and \eqref{horizons} and similar expressions by an appropriate redefinition of the physical charges, but it will reappear in other contributions of electric and magnetic charges e.g. in an accelerated black hole solution considered in \cite{Barrientos:2022bzm}.

\subsection{Cosmological solutions without primordial singularity smoothed by electromagnetic non-linearities}
Many cosmological models, such as  Friedmann-Lemaitre-Robertson-Walker (FLRW) cosmology, face a primordial singularity at the initial time of the development of the universe which indicates that these models (as in the case of the conventional black holes) are not applicable in the region close to the singularity. As was shown in a number of papers (see e.g. \cite{Garcia-Salcedo:2000ujn,DeLorenci:2002mi} and \cite{Novello:2008ra} for a review), non-linear electrodynamics may remove these singularities and give rise to regular cosmological solutions.

The FLRW metric has the following form
\be\label{FLRW}
ds^2=-dt^2+a(t)\left(\frac {dr^2}{1-kr^2}+ r^2\,(d\theta^2+\sin^2\theta\, d\varphi^2)\right),
\ee
where $k=1$ stands for a closed, $k=0$ for flat and $k=-1$ for an open universe. The form of the scalar function $a(t)$ depends on the matter which fills the universe. The FLRW singularity (with infinite density and pressure of the matter) occurs if $a(t_0)=0$ at an initial time $t_0$. This happens e.g. in the case of the FLRW universe filled with a perfect fluid or with Maxwell's electromagnetic fields. However, as was shown in \cite{DeLorenci:2002mi} the singularity can be lifted by non-linear corrections to the Maxwell Lagrangian. For instance, in a simple case of the Heisenberg-Euler-Kockel-like model
\be\label{4thcorrection}
\mathcal L=-\frac 14 F_{\mu\nu}F^{\mu\nu}+\alpha^2 (F_{\mu\nu}F^{\mu\nu})^2+\beta (F_{\mu\nu}\tilde F^{\mu\nu})^2\,,
\ee
where $\alpha$ and $\beta$ are constant parameters, the scalar function $a(t)$ has the following form (for the flat universe)
\be\label{a(t)}
a^2(t)\propto \sqrt{t^2+\alpha^2}
\ee
and is therefore never vanishing.
\if{}
The regularity of this solution is related to the fact that in this example the equation of state for the electromagnetic field energy density and pressure is such that $\rho+3p$, which enters Friedman's equations, becomes negative at small times.
\fi

\subsection{Universe expansion accelerated by NED}
A number of papers studied possibilities of accelerating the expansion of the universe by dominated electromagnetic non-linearities at early time (i.e. inflation) and later stages of its life (see e.g. \cite{Novello:2003kh,Campanelli:2007cg,Kruglov:2015fbl,Ovgun:2017iwg,Kruglov:2020xfc,Kruglov:2020axn} and references therein).  The expansion of the universe is governed by the Freedmann equation for the FLRW scalar $a(t)$
\be\label{a''}
\frac{\ddot{a}}a=-\frac {4\pi}3 (\rho+3p)\,,
\ee
where $\rho$ is the density of a substance that gives a dominant contribution to Einstein's equations and $p$ is its pressure. The universe gets accelerated if $\rho+3p<0$, i.e. when the pressure is negative. As was shown in \cite{Novello:2003kh} this may occur, for instance, if the universe is dominated by non-linear electromagnetic fields described by the following non-linear Lagrangian density
\be\label{1/F}
\mathcal L =S+\frac \beta S, \qquad S=- \frac 14 F_{\mu\nu}F^{\mu\nu}\,,
\ee
where $\beta$ is a constant parameter.

Another model (called rational NED) which produces a similar accelerating effect was considered in \cite{Kruglov:2015fbl}
\be\label{Kruglov}
\mathcal L =\frac {S}{1+2\beta S}\,.
\ee
In these models the values of $\rho$ and $p$ derived from the structure of the electromagnetic energy-momentum tensor are such that $\rho+3p= 2\left(\mathcal L+|\mathbf B|^2\frac{\partial\mathcal L}{\partial S}\right)<0$.

\subsection{NEDs and gravity/CMT holography}
Models of non-linear electrodynamics have also been used for the description of strongly correlated condensed matter systems via gravity/CMT correspondence (see e.g. \cite{Karch:2007pd,Cai:2008in,Myers:2010pk,Jing:2011vz,Baggioli:2016oju,Kiritsis:2016cpm,Blauvelt:2017koq,Kruglov:2018jee,Cremonini:2018kla,Bi:2021maw} and references therein). Gravity/CMT correspondence (see e.g. \cite{Hartnoll:2009sz,zaanen_liu_sun_schalm_2015,Nastase:2017cxp,Baggioli:2021xuv} for a review) is a condensed matter oriented ramification of a powerful conceptual framework called AdS/CFT correspondence, or gauge/gravity duality or holography which has its origin in String Theory (see e.g. \cite{Aharony:1999ti} for a review). This framework provides a toolkit for establishing a dual relation between a weakly coupled gravity theory in an asymptotically anti-de-Sitter space-time and a strongly coupled field theory leaving on the AdS boundary. It prescribes how to translate the physics of the boundary field theory into the gravitational bulk and vice versa, and thus allows to describe strongly coupled non-perturbative quantum systems in terms of the quantities of weakly-coupled gravity, matter and gauge fields.

In applications to condensed matter systems, the holographic set-up in the gravity bulk includes (non-linear) gauge  and matter fields (example, an axion, dilaton and/or charged scalar fields) and is described by the following action
\be\label{graction}
I=\int\,d^4x\,\sqrt{-g}\left(R-2\Lambda+\mathcal L_{NED}(S,P,\phi)+\mathcal L(\phi)\right)+I_{bd},
\ee
where $\mathcal L_{NED}(S,P,\phi)$ is the Lagrangian density of a non-linear electrodynamics coupled to scalar fields, $\mathcal L(\phi)$ is a scalar field Lagrangian density and $I_{bd}$ stands for boundary terms which are required to have a well-defined variational problem. The axion-like fields are usually introduced to break translational invariance in the bulk by taking their profiles be linear in spatial coordinates, which in the boundary CMT corresponds to a momentum dissipation (see e.g. \cite{Baggioli:2021xuv} for a review).

As we have already mentioned, the gravitational background assumed to be an asymptotically AdS black whole whose metric (in many models) has the following form
\be\label{BlackHole}
ds^2=-f\left(r\right)dt^2+\frac{dr^2}{f\left(r\right)}+u\left(r\right)\left(dx^2+dy^2\right)\,,
\ee
where $r$ is the radial holographic coordinate, $x$ and $y$ are associated with the AdS boundary (at $r \rightarrow\infty$), and  $f(r)$ and $u(r)$ are functions of $r$ whose explicit form depends on the structure of the matter and gauge sector of the action \eqref{graction}.  $f(r_h)=0$ determines the black hole horizon.

An often used ansatz for the background electromagnetic field is
\be\label{EMfield}
A_\mu\,dx^\mu=A_t(r) dt+ \frac h2 (xdy-ydx),
\ee
where $h$ is the magnitude of the magnetic field.

The black holes are an important ingredient of the AdS/CMT correspondence since they are thermal states, and their temperature $T=\frac{f'(r_h)}{4\pi}$ is related to the temperature in the dual strongly coupled field theory, while the inclusion of the coupling of a NED to a charged scalar field generalizes that of \cite{Hartnoll:2008vx} for the description of the transition from the normal to the superconducting phase of the superconductor living on the AdS boundary, which in the black hole bulk is manifested by the formation of a scalar field “hair” at and below the critical temperature of the superconductor.

Gravity/CMT holography with NEDs included in the bulk theory was used to study properties of holographic superconductors (see e.g. \cite{Jing:2010zp,Jing:2011vz,Gangopadhyay:2012am} and \cite{Lai:2022tjr} for the latest developments and more references) and for understanding in an analytically controllable way the properties of planar Mott insulators, materials whose insulating character is due to the electron-electron interaction (see \cite{Baggioli:2016oju} and e.g. \cite{Bi:2021maw} for more references). Mott insulators form a class of very important not completely understood condensed matter systems.  An important class of Mott insulators is that of the parent compound of high temperature superconductivity (high-Tc) cuprates with a wide range of applications.

Within the holographic set-up, non-linear electrodynamics has also been used for the description of anomalous scaling laws of strange metals (see \cite{Kiritsis:2016cpm,Blauvelt:2017koq,Cremonini:2018kla} and \cite{Baggioli:2021xuv} for the latest review and references) in which the dynamics of charged degrees of freedom is described by a non-linear Born-Infeld action coupled to gravity and axion scalar fields. In this way the theory reproduces, via non-linear electromagnetic interactions, the anomalous temperature dependence of the resistivity $\rho \sim T$ and Hall angle $\cot\theta_H \sim T^2$ of the cuprate strange metals.\footnote{These materials are called strange because their behavior deviates from the standard properties of metals. In particular, they are characterized by an anomalous temperature dependence of resistivity. While the resistivity of the conventional metals increases with the square of the temperature, the electrical resistivity of a wide class of strange metals increases linearly with temperature. }

A basic idea behind the holographic description of these materials is that the electric current operator $j^\mu$ describing the charged sector is holographically dual to the electromagnetic field $A_\mu$ in the $AdS$  black hole bulk. So, current-current interactions in the condensed matter system are associated with self-interactions of the electromagnetic field via non-linear terms of $F_{\mu\nu}$ in the NED Lagrangian.
The physical characteristics of the system depend, of course, on the choice of the concrete model of electrodynamics. For instance, for a wide class of models, in the probe field limit, the components of the electrical conductivity matrix are given by \cite{Iqbal:2008by,Guo:2017bru}
$$
\sigma_{xx}=\sigma_{yy}=\frac{\partial \mathcal L\left(S,P,\phi\right)}{\partial S}|_{r_h}, \qquad  \sigma_{xy}=-\sigma_{yx}=- \frac{\partial \mathcal L\left(S,P,\phi\right)}{\partial P}|_{r_h}
$$
which are evaluated at the black brane horizon $r=r_h$.

Using these relations one can compute e.g. the electric resistivity and the Hall angle of the material
\be\label{RH}
\rho_{xx}=\frac{\sigma_{xx}}{\sigma^2_{xx}+\sigma^2_{yy}}\,, \qquad \cot\Theta_H =\frac{\sigma_{xx}}{\sigma_{yy}}\,,
\ee
 and study their properties in relation to temperature, phase transitions etc. by applying the holography tools.

\section{Birefringence in non-linear electrodynamics}
Let us now discuss a physical phenomenon which is common to almost all models of non-linear electrodynamics and which can be tested by an experiment.\footnote{Another phenomenon associated with electromagnetic non-linearities of the QED is light-by-light scattering which was only quite recently observed experimentally \cite{ATLAS:2017fur,ATLAS:2019azn}.} This is the birefringence of light propagating in a uniform strong electromagnetic background.  Birefringence is a phenomenon of double refraction such that a ray of light is split by polarization, with respect to the optical axis of the material, into two rays taking different geodesic paths. In the NED, the electromagnetic background plays the role of an optical material.

To compute the birefringence effect in non-linear electrodynamics one splits the electromagnetic field into the background field (satisfying the NED equations of motion) and small fluctuations around this background $F_{\mu\nu}=F^{ext}_{\mu\nu}+f_{\mu\nu}$, derives the form of the Lagrangian up-to the second order in the fluctuations and computes the dispersion relations determining the propagation of plane waves in a stong electromagnetic background \cite{Boillat:1970gw,Plebanski:1970zz,Bialynicka-Birula:1970nlh}. In general, the dispersion relations depend on the orientation of the polarization of the propagating light (its electric field direction) with respect to the axis of the background field. Let us assume that the background field is purely magnetic $\mathbf B_{ext}$. Then the dispersion relations have the following form
\be\label{dispersion}
(\parallel) \quad \omega^2=\frac{|{\bf k}|^2}{n^2_{\parallel}}\,\,,\qquad (\perp) \quad \omega^2=\frac{|{\bf k}|^2}{n_\perp ^2}\,,
\ee
where $n_{\parallel}(\mathbf k,\mathbf B_{ext})$ and $n_{\perp}(\mathbf k,\mathbf B_{ext})$ are refraction indices. Their subscripts indicate that the electric field of the propagating linearly polarized light is parallel or orthogonal to $\mathbf B_{ext}$. The values of the refraction indices depend on properties of the NED and the background. For instance,
\begin{itemize}
\item
in  Maxwell's theory $n_{\parallel}=n_{\perp}=1$, the light propagates along a relativistic geodesic and there is no birefringence.
\item
The birefringence is also absent in the Born-Infeld theory \eqref{BI} in which $n_{\parallel}=n_{\perp}$ and (for $\mathbf k \perp \mathbf B_{ext}$)
\be\label{BId}
\omega^2=|\mathbf k|^2\left(1+\frac{|\mathbf B_{ext}|^2} T\right)^{-1}\,.
\ee
One can see that the speed of light in the Born-Infeld electromagnetic background is less than the speed of light in the vacuum. So there is no problem with acasual propagation.
 The Born-Infeld theory is the unique (physically consistent) non-linear electrodynamics which has no birefringence \cite{Boillat:1970gw,Plebanski:1970zz}.
\item
In the Heisenberg-Euler-Kockel effective theory of the spinor QED (for $\mathbf k \perp \mathbf B_{ext}$)
\be\label{HEKb}
(\parallel ) \quad \omega^2=|{\bf k}|^2\left(1-14 \frac {2\alpha^2}{45 m_e^4}|\mathbf B_{ext}|^2\right)\,,\qquad  (\perp)\quad  \omega^2=|{\bf k}|^2\left(1-8 \frac {2\alpha^2}{45 m_e^4}|\mathbf B_{ext}|^2\right)\,,
\ee
and the magnitude of birefringence $(\Delta n=n_{\parallel}-n_{\perp})$
\be\label{deltan}
\Delta n_{QED}=3.96 \times 10^{-24} |\mathbf B_{ext}|^2\,({\rm Tesla})^{-2}\,,
\ee
depends on the magnitude of the magnetic field.
\item
In the ModMax theory the dispersion relations are \cite{Bandos:2020jsw,Flores-Alfonso:2020euz}
\be\label{ModMaxdispersion(i)}
(\perp) \quad n_\perp=1,\qquad  \omega^2=|\bf k|^2 \,,
\ee
which is the standar relativistic light-cone, and another one 
\be\label{ModMaxdispersion(ii)}
\omega^2 = |{\bf k}|^2 (\cos^2 \varphi + e^{-2\gamma}\sin^2\varphi),
\ee
where $\varphi$ is the angle between ${\bf k}$ and $\bf B_{ext}$\,. Note that in ModMax, in contrast to the Born-Infeld and Heisenberg-Euler theory, the dispersion relations do not depend on the magnitude of the background field $|\mathbf B_{ext}|$ but only on the angle between ${\bf k}$ and $\bf B_{ext}$ and the coupling constant $\gamma$. This is a consequence of conformal invariance. The propagation of this mode is always sub-luminal if $\gamma > 0$. As in all other cases of non-linear electrodynamics, the effect of the magnetic background is maximal when the light propagates in a direction orthogonal to $\bf B_{ext}$ ($\varphi=\frac \pi 2$) while its electric component is parallel to $\bf B_{ext}$:
\be\label{kB}
(\parallel)\quad  \omega^2 = |{\bf k}|^2 e^{-2\gamma}\,, \qquad n_\parallel =e^\gamma\,.
\ee
For the rays with the dispersion relations \eqref{ModMaxdispersion(i)} and \eqref{kB} the magnitude of birefringence is
\be\label{deltaMM}
\Delta n_{ModMax}=e^\gamma-1\simeq \gamma+O(\gamma^2)\,.
\ee
\end{itemize}
Electromagnetic vacuum birefringence has been searched for in a number of experiments one of which is PVLAS \cite{Ejlli:2020yhk}. The experiment was not able to reach the predicted value from QED \eqref{deltan}, but set the current best limit on vacuum birefringence in an external magnetic field $\mathbf B_{ext}$ = 2.5 Tesla\footnote{An indirect evidence associated with vacuum birefringence of light passing through strong magnetic fields generated by a neutron star was reported in \cite{Mignani:2016fwz}. Another indirect evidence was announced recently by the STAR collaboration at RHIC \cite{STAR:2019wlg}.}
\be\label{deltaPVLAS}
\Delta n_{exp}\leq (12 \pm 17)\times 10^{-23}\,\,\, {\rm for}\,\,\, |\mathbf B_{ext}|=2.5\,\,{\rm Tesla}\,.
\ee
A new PVALS experiment which should come in operation in a near future will use more powerful magnets
 % $|\mathbf B|_{ext}\sim 8\, \rm T$
which will allow to reach the QED value of birefringence.

The PVLAS experimental bound \eqref{deltaPVLAS} allows to estimate a bound on the value of the coupling constant $\gamma$ of the ModMax electrodynamics which (as follows from \eqref{deltaMM}) is
$$
\gamma \leq 3 \cdot 10^{-22}\,.
$$
So if ModMax has something to do with fundamental physics this raises the question why its coupling constant is so small. This of course concerns the parameters of any NED. On the other hand, in the effective field theory applications of NEDs to the description of condensed matter systems their parameters may be large.

One should however note that the birefringence in external electromagnetic background may be a combined effect of different classical and quantum non-linear contributions which may either enhance or decrease it (like in the $N=1$ supersymmetric non-linear electrodynamics in which $\Delta n=0$ for the non-linearities of the fourth order in $F_{\mu\nu}$). So it is important to have precise experimental results on the values of $n_{\parallel}$ and $n_\perp$ for being able to falsify one or another NED.

To conclude this section let me note that NED has also been considered in the context of cosmic birefringence that has long been regarded as a promising avenue for new physics. A particularly important scenario is Cosmic Microwave Background (CMB) birefringence induced by the coupling between a (pseudo)scalar axion-like field and the electromagnetism, which causes a rotation of the linear polarization plane of CMB photons during propagation \cite{Planck:2016frx}. Cosmological birefringence has nowadays attracted a renewed interest (see e.g. \cite{Greco:2022ufo} and references therein), following its claimed extraction from Planck 2018 data  \cite{Minami:2020odp}. Cosmic birefringence can also manifest itself due to Euler-Heisenberg photon-photon forward scattering that produces circular polarized CMB photons \cite{Inomata:2018vbu,Hoseinpour:2020hic} for which new observational constraints have recently been obtained \cite{SPIDER:2017kwu,CLASS:2019bgh}.

For other measurable effects which can be used to test the NEDs see e.g. \cite{Fouche:2016qqj,NiauAkmansoy:2017kbw,Neves:2021tbt,DeFabritiis:2021qib}.

\section{Outlook}
For almost 90 years non-linear electrodynamics has been an active area of research in theoretical physics with applications ranging from the theory of fundamental interactions and cosmology to condensed matter systems and non-linear optics. All (or most of) the models of non-linear electrodynamics should be regarded as effective field theories. Only for few of them, namely the Born-Infeld and Heisenberg-Euler electrodynamics, we know the corresponding fundamental theories, such as String Theory and QED. As was discussed in \cite{Gibbons:2000xe}, some NEDs may arise upon a Kaluza-Klein dimensional reduction of a multi-dimensional higher-curvature gravity, which in turn can be a low energy limit of String Theory. It would be of interest to look for a fundamental origin of other NEDs, in particular, of ModMax whose structure is uniquely determined by two fundamental symmetries, electric-magnetic duality and conformal invariance. Recently, Nastase \cite{Nastase:2021uvc} (see also \cite{Ferko:2022iru}) discussed a possibility of interpreting the generalized Born-Infeld electrodynamics, which reduce to ModMax in the weak field limit, in terms of a brane-like construction within a string theory. An origin of ModMax and its Born-Infeld generalization from a $T\bar T$-like deformation of Maxwell's theory has been revealed in \cite{Babaei-Aghbolagh:2022uij,Ferko:2022iru,Conti:2022egv}.
One may also wonder whether ModMax arises as a certain (conformal) limit of a theory in which electromagnetic fields interact with some matter fields (e.g. axion-dilaton-like) and the latter are integrated out in the effective action. A related interesting question is whether ModMax can be consistently quantized.

Work on these and other NED issues and their applications is going ahead.

\section*{Acknowledgements}
I am grateful to Kurt Lechner for useful comments on these notes.
This work has been partially supported by the INFN Research Project ``String Theory and Fundamental Interactions''
(STEFI), by the Spanish MICINN/FEDER (ERDF EU) grant PGC2018-095205-B-I00 and by the Basque Government Grant IT-979-16.

%\newpage

\if{}
\bibliographystyle{abe}
\bibliography{references}{}
\fi

\providecommand{\href}[2]{#2}\begingroup\raggedright\endgroup

\end{document}